\documentclass[twocolumn,amsmath,amssymb,aps,floatfix, longbibliography,superscriptaddress]{revtex4-1}
\usepackage{graphicx}
\usepackage{dcolumn}
\usepackage{bm}
\usepackage{amssymb}

\begin{document}

\title{Oscillation in the temperature profile of the large-scale circulation of turbulent convection induced by a cubic container} 
\author{Dandan Ji}
\affiliation{Department of Physics, Yale University, New Haven, CT 06520}
\author{Eric Brown}
\affiliation{Department of Mechanical Engineering and Materials Science, Yale University, New Haven, CT 06520}
\email{ericmichealbrown@gmail.com}

\date{\today}

\begin{abstract}
We present observations of oscillations in the  shape of the temperature profile of the large-scale circulation (LSC) of turbulent Rayleigh-B{\'e}nard convection.   Temperature measurements are broken down into Fourier moments as a function of $\theta-\theta_0$, where $\theta$ is the azimuthal angle in a horizontal plane at mid-height, and $\theta_0$ is the LSC orientation.  The oscillation structure is dominated by a 3rd order sine moment and 3rd order cosine moment in a cubic cell.  In contrast, these moments are not found to oscillate in a cylindrical cell.  This geometry-dependent behavior can be explained by a minimal model that assumes that the heat transported by the LSC is conducted from the thermal boundary layers, and is proportional to the pathlength of the LSC along boundary layers at the top and bottom plates.  In a non-circular cross-section cell, oscillations of the LSC orientation $\theta_0$ result in an oscillation in the container shape in the reference frame of the LSC, resulting in an oscillation in the pathlength of the LSC at a given $\theta-\theta_0$.  In a square-cross-section cell, this model predicts the dominant 3rd order sine moment and 3rd order cosine moment with magnitudes within 50\% of measured values, when using the amplitude of the oscillation of $\theta_0$ as input.   A cylindrical cell is special in that the pathlength is independent of $\theta_0$, and so these oscillating moments are not induced.  In a cylindrical cell, the model reproduces the sinusoidal mean temperature profile with a sloshing oscillation dominated by the 2nd order sine moment, consistent with previous observations in that geometry.
\end{abstract}

\maketitle

%PhySH: fluid dynamics: RBC, turbulence, stochastic differential equations

\section{Introduction}

%motivation -LSC & dynamics
Turbulent flow often generates large-scale coherent structures, such as convection rolls in the atmosphere.   Such structures can play important roles in directing heat and mass transport. The fact that these structures have typical shapes and are long-lived makes them more amenable to being approximately modeled by equations of motion, more so than the wide spectrum of small-scale turbulent fluctuations.  A long-term goal is more complete modeling of such large-scale coherent flows in turbulence.  We work toward this goal by presenting a model to explain a recently observed oscillation in the shape of the temperature profile of a large-scale coherent structure, described below. 

%\subsection{Model system}
We use Rayleigh-B\'enard convection as a model laboratory system that generates large-scale coherent flow structures (for reviews, see Refs.~\cite{AGL09,LX10}).  A  fluid is heated from below and cooled from above to generate buoyancy-driven convection. This system exhibits robust large-scale coherent structures, i.e.~they retain a particular organized flow structure over long times.  For example, in containers of aspect ratio near 1, typically a large-scale circulation (LSC) forms. This LSC consists of  temperature and velocity fluctuations which, when coarse-grain averaged, collectively form a single convection roll in a vertical plane that breaks the symmetry of the system, and exists nearly all of the time \cite{KH81}. Turbulent fluctuations drive the orientation $\theta_0$ of the LSC in the horizontal plane to meander spontaneously and erratically over time  \cite{CCS96, BA06a, XZX06}.  %Many dynamics of the LSC can be described by a model of diffusive motion (representing turbulent fluctuations) in potentials corresponding to the various forces tending to change the strength and orientation of the LSC \cite{BA08a,BA08b, AAG11, SBHT14, BJB16,SLZ16,  ZLW17,  JB20a}.

%n=2 oscillation modes
The dynamics of the LSC includes a prominent oscillation mode that has long been observed in leveled cylindrical containers  \cite{HCL87, SWL89, CGHKLTWZZ89, CCL96, TSGS96, CCS97, QYT00, QT01b, NSSD01, QT02, QSTX04, FA04, SXT05,TMMS05}.   The flow  structure of these oscillations is a combination of twisting in $\theta_0$ \cite{FA04, FBA08} and  sloshing \cite{XZZCX09, ZXZSX09, BA09}.    This combination of flow structures can alternatively be described from a Lagrangian viewpoint as a single closed-loop advected oscillation mode, consisting of the hot and cold regions oscillating in the azimuthal coordinate while being advected around by the LSC with two oscillation periods per LSC rotation (referred to as the $n=2$ advected mode) \cite{BA09}.  This oscillation can be modeled with stochastic differential equations \cite{BA09}.  The pressure from the sidewall creates a potential, with a corresponding restoring force against slosh displacements which pushes the LSC towards alignment with the center of the container  \cite{BA09}.  Turbulent fluctuations provide a broad-spectrum noise to drive oscillations at the resonant frequency  of the potential.    

%non-circular n=1
In non-circular cross-section containers, the potential is further modified by the geometry of the cell, and the restoring force tends to push the LSC to align with the longest diagonals \cite{Zimin1978,ZML90,QX98, VPCG07, JB20a}.  This results in oscillations in $\theta_0$ around a potential minimum at a corner in rectangular cross-section containers \cite{SBHT14, VSFBFBR16, GKKS18, JB20a}.  The addition of a restoring force acting directly on $\theta_0$ excites additional advected modes.  In this case, the dominant structure has one oscillation per LSC rotation (referred to as the $n=1$ advected oscillation mode), in which the oscillation in $\theta_0$ is in-phase at different heights, and the sloshing is out-of-phase at the top and bottom halves of the container \cite{BA09,JB20a}. 
 
%other oscillation modes
Several other oscillation modes are known to exist.   The LSC orientation $\theta_0$ can oscillate between diagonals of a non-square rectangular cross-section in a wider potential well \cite{SBHT14}.   In cylindrical cells tilted by several degrees relative to gravity, the LSC orientation can oscillate around the orientation of the tilt; the tilt produces an orientation-dependent forcing that provides a restoring force that can cause an oscillation  around a potential minimum \cite{BA08b}.   An oscillation in $\theta_0$ was found in a cubic cell where the preferred orientation aligns with grooves in the top and bottom plates \cite{FNAS19}, which can also be described as an oscillation around a preferred orientation.     An oscillation mode in which the rotational axis of of the LSC flexes and rotates around the center of the cell like a jump rope has been observed in aspect ratio 2 cells, although its cause is unknown \cite{VHGA18}. 
%VSFBFBR16 also observed a weaker oscillation in temperature/vertical component of flow \cite{Frick}

%motivating work
In a previous manuscript, we reported observations of a new oscillation in the shape of the LSC \cite{JB20a}.  This was represented in terms of Fourier moments of the temperature profile in  the azimuthal coordinate relative to the orientation $\theta_0$ of the LSC \cite{JB20a}.  Specifically, we found 3rd- and 4th-order sine moments in a cubic cell, which were not found in cylindrical cells \cite{BA09}, suggesting that this oscillation depends on cell geometry.    The 3rd-order moment in the temperature profile does not have the antisymmetry of the temperature profile of the sloshing mode (where the hot and cold sides of the temperature profile move in opposite directions) \cite{BA09}, or the symmetry around the LSC plane of the jumprope oscillation \cite{VHGA18}, so is distinct from those known modes.   As the goal of those measurements was to test the prediction of advected oscillation modes in a cubic container, which required measuring the sloshing oscillation but not the full temperature profile, an incomplete set of moments consisting of only sine terms was presented.  %The other known oscillation modes correspond to oscillations of the LSC orientation, and other oscillations in the shape of the temperature profile have been reported.

%In this manuscript...
In this manuscript, we present a more complete set of Fourier moments of the temperature profile, and we present a minimal model to explain the oscillation in the shape of the temperature profile of the LSC in a non-circular cross-section container \cite{JB20a}.  Section \ref{sec:methods} explains the experimental methods.  Section \ref{sec:cube_structure} shows the observations of the temperature profile of the LSC in a cubic cell in terms of Fourier moments of the temperature profile, reporting  a complete set of both sine and cosine moments.  Section \ref{sec:cylinder_structure} compares these to the same Fourier moments for a cylindrical cell to confirm that the new moments are not found in cylinders. Section \ref{sec:model} presents a model to explain the oscillation in the temperature profile.

\section{The experimental apparatus and methods}
\label{sec:methods}

\subsection{Experimental setup}
\begin{figure}
\includegraphics[width=0.475\textwidth]{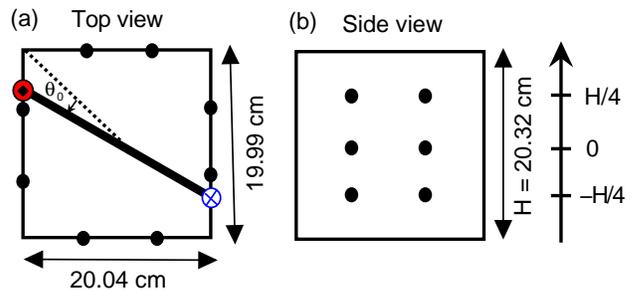} 
\caption{A schematic of the  experimental setup. (a) Top view: the thick line represents the central circulation plane of the LSC, with $\theta_0$ defined as the orientation relative to a diagonal.   Thermistor locations on the sidewall are indicated by solid circles.  (b) Side view.  
}
\label{fig:setup}
\end{figure}   

The design of the  experimental setup is the same apparatus used in Ref.~\cite{BJB16}, where it is described in more detail. We describe it briefly here. The  flow chamber is nearly cubic, with height  $H = 20.32$ cm, and horizontal dimensions of $20.04$ cm and $ 19.99$ cm, as shown in Fig.~\ref{fig:setup}.  The top and bottom surface are aluminum plates, with double-spiral water-cooling channels.   To  control the plates at a temperature difference of $\Delta T$, with the bottom plate  hotter than the top plate, water is circulated through each plate by a temperature-controlled water bath. The sidewalls are made of plexiglas, and three out of four have a thickness of 0.55 cm.  The fourth sidewall has thickness $0.90$ cm, and is shared with another identical  flow chamber to be used in future experiments, but the two  chambers are designed to be thermally insulated from each other for the experiments reported here. 
The working fluid is degassed and deionized water with mean temperature of 23.0$^{\circ}\mathrm{C}$, for a Prandtl number $Pr = \nu/\kappa = 6.4$, where $\nu=9.4\times10^{-7}$ m$^2$/s is the kinematic viscosity and $\kappa$ is the thermal diffusivity. The Rayleigh number is given by $Ra= g\alpha \Delta T H^3/\kappa\nu$ where $g$ is the acceleration  of gravity and $\alpha$ is the thermal expansion coefficient.  We report measurements at $\Delta T=18.35$ $^{\circ}$C  for $Ra = 2.62\times10^9$.
% refer to RBC_cali_July2015 for more info
%kappa=1.24x10^-3 cm^2/s

%alignment
  The alignment of the LSC is partially locked along a diagonal to prevent corner-switching with a small tilt of the cell relative to gravity, as is commonly done \cite{KH81}.  We report data for a tilt angle of $\beta = (1.0\pm0.2)^{\circ}$, at an orientation within $0.03$ rad of the diagonal where we define $\theta_0=0$.   
 
%orientation locking
 %All of the aforementioned variations away from a perfectly cubic cell can introduce asymmetries that in principle can affect the flow dynamics \cite{BA06b}.  However, we will see in Sec.~\ref{sec:potential} that the cubic shape is the dominant geometric factor, in particular as the preferred orientation of the LSC tends to align closely along the diagonals of the cell.  Which of the diagonals chosen may be affected by the  aforementioned asymmetries, and spontaneous switching between diagonals can occur \cite{BJB16,FNAS17, GKKS18}.

   To measure the LSC, thermistors were mounted in the sidewalls.   There are 3 rows thermistors at heights $-H/4$, $0$, and $H/4$  relative to the mid-height of the container, as shown in Fig.~\ref{fig:setup}b.   In each row there are 8 thermistors equally spaced in the azimuthal angle $\theta$  measured in  a horizontal plane, as shown in Fig.~\ref{fig:setup}a.     In this manuscript, we only report measurements from the middle row of thermistors, unless otherwise stated. Temperatures were recorded every 9.7 s at each thermistor.

  \subsection{Obtaining the LSC orientation $\theta_0$}
\label{sec:lsc}

 The LSC orientation was obtained using the same methods as Ref.~\cite{BA06a}.  As the LSC pulls hot fluid from near the bottom plate up one side and cold fluid from near the top plate down the other side, a temperature difference is detected along  a horizontal direction at the mid-height of the container.  We fit the 8 thermistor temperatures at the middle row to the function 
\begin{equation}
T = T_0 + \delta cos(\theta - \theta_0)
\label{eqn:delta}
\end{equation}
to get the orientation $\theta_0$ and strength $\delta$ of the LSC at each time step. 
%  we have a systematic error of 4.0 mK, random error of 0.7 mK on the temperature measurements. Propagating the relative error on the thermistor measurements \cite{JB20a} leads to the uncertainties on $\theta_0$ of  $\delta T/\delta\sqrt{8} =1.5/\delta$ mK/rad and  $\delta$ of $\delta T/\sqrt{8} = 1.5$ mK, corresponding to 0.003 rad and 0.3\% errors, respectively, at our typical Ra $=2.62\times10^9$. Another random error comes from the fit, which is on average 12\% of $\delta$, and 0.12 rad on $\theta_0$ at Ra$=2.62\times10^9$. 

\subsection{Obtaining the Fourier moments of the temperature profile}
\label{sec:Fouriermethods}

 We first fit Eq.~\ref{eqn:delta} to $T(\theta)$ to obtain $\theta_0$ at each time step.  This allows viewing of the temperature profile relative to the LSC orientation as $T(\theta-\theta_0)$ to separate out meandering and oscillations of the LSC orientation from changes in the temperature profile.  We then express perturbations on the sinusoidal temperature profile as the Fourier series 

\begin{multline}
T = T_0 + \delta \cos(\theta - \theta_0) + \sum_{n = 2}^{4} A_n \sin(n(\theta - \theta_0)) \\+ \sum_{n = 2}^{4} B_n \cos(n(\theta - \theta_0)) \ ,
\end{multline}
 where the Fourier moments are calculated as
\begin{equation}
A_n = \frac{1}{4}\sum_{i = 1}^{8}{[T_i - T_0 - \delta \cos(\theta - \theta_0)]\sin[n(\theta - \theta_0)]}
\label{eqn:a_n}
\end{equation}
and 
\begin{equation}
B_n = \frac{1}{4} \sum_{i = 1}^{8}{[T_i - T_0 - \delta \cos(\theta - \theta_0)]\cos[n(\theta - \theta_0)]} \ . 
\label{eqn:b_n}
\end{equation}
Here the sum over $i$ corresponds to the sum over different thermistors.  We can only calculate the first 4 moments due to the Nyquist limit with 8 measured temperatures.  We don't include $B_1$ since that was already defined as $\delta$, and $A_1$ is trivially zero since we first fit to Eq.~\ref{eqn:delta}.

\section{Fourier moments of the temperature profile of the LSC in a cubic cell}
\label{sec:cube_structure}
 \label{sec:tempprofile}

%shape oscillations
In this section, we identify oscillations in the temperature profile of the LSC in a cube in terms of Fourier moments of the temperature profile $T(\theta-\theta_0)$ as in Refs.~\cite{BA09, VHGA18}. 
\begin{figure}
\includegraphics[width=.475\textwidth]{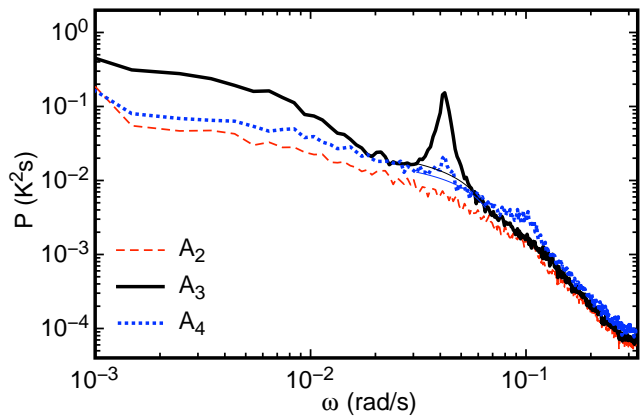} 
\caption{Power spectrum of the Fourier moments $A_n$, as indicated in the legend.    Thin lines show the linear background $P_{bg}$ subtracted off when integrating the peak power.    Each moment oscillates, except for $A_2$.  
}
\label{fig:power_spec_momentsA}
\end{figure}

\begin{figure}
\includegraphics[width=.475\textwidth]{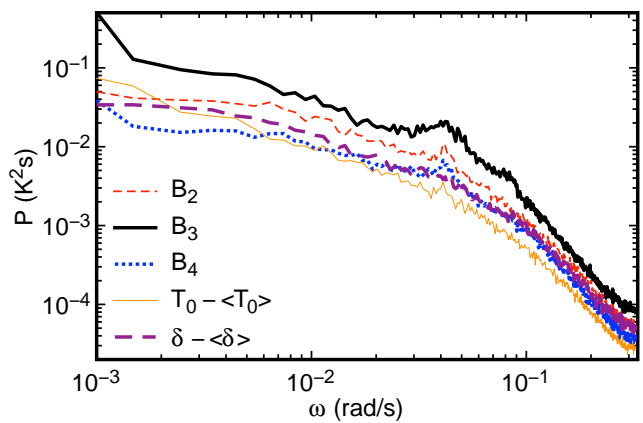} 
\caption{ Power spectrum of the Fourier moments $B_n$ and other terms of the temperature profile, as indicated in the legend.  Each signal  oscillates except for $\delta-\langle\delta\rangle$. 
}
\label{fig:power_spec_momentsB}
\end{figure}

% data from 151123_20C longrun
\begin{figure}
\includegraphics[width=.475\textwidth]{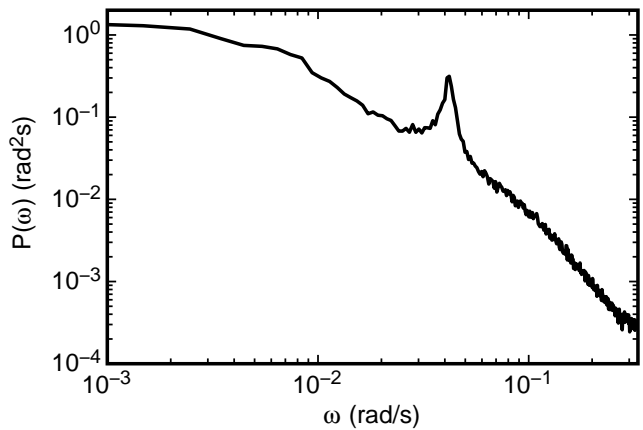} 
\caption{Power spectrum of $\theta_0-\langle\theta_0\rangle$.  The oscillations of the Fourier moments have the same frequency as the oscillation of $\theta_0$.}
\label{fig:power_spec_theta_mid}
\end{figure} 

% power spec 
Fourier moments were calculated according to Sec.~\ref{sec:Fouriermethods}, and the power spectrum of each Fourier moment was calculated as 
\begin{equation}
P(\omega) = \frac{1}{N\pi}\sum \left| \hat x(\omega)\right|^2 dt
\label{eqn:powerspec}
\end{equation}
where $N$ is the number of data points, and $\hat x(\omega) = \sum x(t) \exp(i\omega t)$ where $x$ can stand for any parameter of the temperature profile such as $A_n$, $B_n$, $\delta-\langle\delta\rangle$, or $T_0-\langle T_0\rangle$.  In each case we average $P(\omega)$ over a range of $9.88\times10^{-4}$ rad/s in $\omega$ to smooth the data.  The smoothed $P(\omega)$  are shown in Fig.~\ref{fig:power_spec_momentsA} for $A_n$ ($A_n$ are reproduced from \cite{JB20a}) and in Fig.~\ref{fig:power_spec_momentsB} for $B_n$, $\delta-\langle\delta\rangle$, and $T_0-\langle T_0\rangle$.  We also show the power spectrum of $\theta_0-\langle\theta_0\rangle$ in Fig.~\ref{fig:power_spec_theta_mid} for comparison. Most of the power spectra have peaks ($T_0-\langle T_0\rangle$, $B_2$, $B_3$, $B_4$, $A_3$, and $A_4$), indicating oscillation, which is at the same frequency as the oscillation of $\theta_0$.
% Parseval's theorem satisfied for power of whole dataset within 5\%.  see page 93 of notebook
%  frequency is $0.042 \pm 0.001$ rad/s --
Only moments $A_2$ and $\delta-\langle\delta\rangle$ do not have a resolvable peak in the power spectrum.  

\begin{figure}
\includegraphics[width=.475\textwidth]{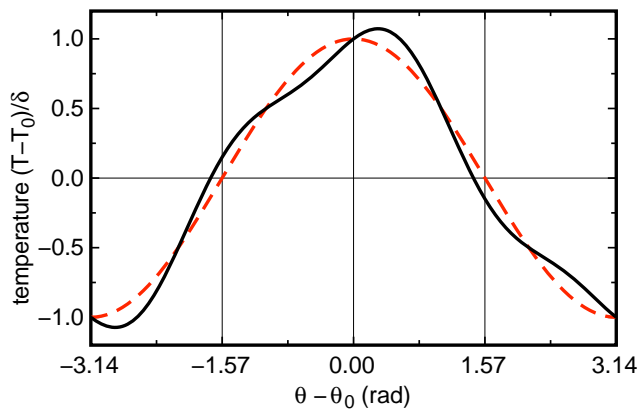} 
\caption{Solid line: Illustration of a temperature profile with an  $A_3$ moment with amplitude of $0.15\delta$. Dotted line: $\cos(\theta-\theta_0)$.
}
\label{fig:A3mode}
\end{figure}

%n=1 mode
Some of the Fourier moments are already known as contributing to advected oscillation modes \cite{BA09,JB20a}. The oscillation in $\theta_0$ is part of the $n=1$ advected oscillation mode which was predicted and observed to be dominant in cubic containers \cite{JB20a}, in which the oscillation in $\theta_0$ has maximum amplitude at the mid-height of the container, and nodes at the center of the top and bottom plates \cite{BA09,JB20a}.  The sloshing, typically represented by the moment $A_2$, is predicted to have nodes at the mid-height of the container, and maximum amplitude at the center of the top and bottom plates \cite{BA09,JB20a}.  The lack of an $A_2$ moment at the lowest frequency at mid-height in a cubic geometry is thus predicted by the advected oscillation model.

%interpretation of higher order moments A_n and B_n
Other Fourier moments of the temperature profile can be related to oscillation structures as follows. In principal, any even-order $A_n$ (i.e.~$A_2$ and $A_4$)  shift the maximum and minimum of the temperature profile in opposite directions, are antisymmetric around $\theta_0$, and could be interpreted as a sloshing of the LSC \cite{XZZCX09, ZXZSX09, BA09}.  Only the $A_2$ moment was found to oscillate in cylindrical containers, and $A_4$ has not been reported to oscillate before \cite{BA09}.  Even-order $B_n$ (i.e.~$B_2$ and $B_4$) and $T_0-\langle T_0\rangle$ lead to a change in the temperature profile that breaks the symmetry around $T=\langle T_0 \rangle$, and can split one of the temperature extrema into two separate extrema if $B_2$ or $B_4$ are large enough, similar to the jumprope oscillation \cite{VHGA18}.  Our observation of small peaks in $B_2$, $B_4$, and $T_0-\langle T_0\rangle$ confirm the existence of a weak jumprope mode, or a mode with similar symmetry for water in a cubic cell.    Odd-order $B_n$ (i.e.~$B_3$)  preserve all the symmetries of the temperature profile, but have not been reported to oscillate before.  Odd-order $A_n$ (i.e.~$A_3$) shift the maximum and minimum of the temperature profile in the same direction by the same amount, but break the symmetry around $\theta_0=0$, while preserving a translational antisymmetry upon shifting $\theta_0$ by $\pi$.  An illustration of this moment is given in Fig.~\ref{fig:A3mode}. The $A_3$ moment does not correspond to any previously reported oscillation modes.

%140916
%A3 is consistently pi out of phase with $\theta_0$ for all 4 wells

%peak power
\begin{table}
\begin{tabular}{rrrrrrrrrr}
%\begin{tabular}{|r|r|r|r|r|r|r|r|r|r|}
 \hline\hline
 moment  & $A_{2,p}$ & $A_{3,p}$ & $A_{4,p }$ & $B_{2,p}$ & $B_{3,p}$  & $B_{4,p}$ & $T_p$ & $\delta_p$ &   $\theta_{0,p}$\\ 
 \hline
amplitude (mK)  & 0 & 28   & 7 &  3& 7  & 2 & 2 & 0 & 16  \\ 
 \hline\hline
\end{tabular} 
\caption{Root-mean-square oscillation amplitudes of terms of the temperature profile.  All amplitude are given in units of mK, each with an error of 2 mK.  The amplitude of the oscillation in  $A_{3}$ is significantly larger than the other terms. 
}
\label{tab:peak_power}
\end{table}

%quantify power spectrum
To quantify the amplitude of each of the oscillating moments in  Figs.~\ref{fig:power_spec_momentsA}, \ref{fig:power_spec_momentsB}, and \ref{fig:power_spec_theta_mid}, we calculate the peak power in each power spectrum.  Each peak power is obtained from the integral $\int (P(\omega)-P_{bg})d\omega$, where $P_{bg}$ is a linear background which is excluded from the integral, as shown for example in Fig.~\ref{fig:power_spec_momentsA} as the region bounded by the thin line that connects the first and last data point in the visually identified peak.   According to Parseval's Theorem, the  root-mean-square (rms) amplitude of the oscillation in units of temperature is given by the square root of the integral of peak power.  We define these rms oscillation amplitudes as $A_{n,p}$ for moments $A_n$,  $B_{n,p}$ for moments $B_n$,  $\delta_p$ for $\delta-\langle\delta\rangle$, and $T_p$ for $T_0-\langle T_0\rangle$.  These oscillation amplitudes are summarized in Table \ref{tab:peak_power}. There is an error of 2 mK on these measurements, dominated by the uncertainty in identifying the range to fit the peak and its background.  The raw amplitude of $\theta_{0,p}=37$ mrad is obtained from the power spectrum of $\theta_0-\langle\theta_0\rangle$, which is converted into mK for comparison in Table \ref{tab:peak_power} by multiplying by $\langle\delta\rangle = 0.431$ K.    Among all the amplitudes,   $A_{3,p}$ is the largest, even larger than the well-known oscillation in $\theta_0$, and $B_{3,p}$ and $A_{4,p}$ are tied for the next largest amplitudes.  The remaining moments are near the resolution limit.  The significant oscillations in the moments $A_3$, $B_3$, and $A_4$ remain unexplained.
% The background of the power spectrum of $A_3$ is 101 mK, indicating that even the strongest oscillation amplitudes are weak compared to the total power of fluctuations in the turbulent system. 
%Secondary peaks can be seen for moments $A_n$ and $B_3$ near  $\omega = 0.1$.
% background over identified peak frequency range  is 7 mK

\subsection{Asymmetries}

%$A_4$ mode
To check for consistency of data with expected symmetries, we measured phase shifts between different oscillating moments by calculating correlation functions (not shown) as in \cite{JB20a}. Notably, the oscillation of moment $A_4$ is found to be in phase with $\theta_0$.  % and out of phase with $A_3$ %151123
 An in-phase superposition of these two moments results in an asymmetry between the hot and cold sides of the LSC at the mid-height where the hot side oscillates with a larger amplitude than the cold side.   This asymmetry between the hot and cold sides of the flow appears to be a form of non-Boussinesq asymmetry, similar to an observation of a very weak oscillation of $A_2$ out-of-phase by $\pi$ rad with $\theta_0$ \cite{JB20a}.  To test whether this is a non-Boussinesq effect due to the change in material parameter values with the temperature on opposite sides of the cell, or due to some other asymmetry of the setup, we compare data from a similar dataset where the cell was nearly level but tilted slightly to cancel out asymmetries so that the LSC orientation sampled all 4 corners of the container \cite{BJB16}.   When calculating the phase shift between $A_4$ and $\theta_0$ separately for each potential well, we find them to be in phase in 2 wells and $\pi$ out of phase in the other 2 wells. This difference between potential wells suggests that the existence of $A_4$ in the long-term average is likely due to an asymmetry of the experimental setup.  An alternative possibility is that it is a long-lived spontaneous-symmetry-breaking mode that did not have time to sample all of the possible phase shifts, although it was observed in a dataset that was 10 days long.

\section{Fourier moments of the temperature profile in a  cylindrical cell}
\label{sec:cylinder_structure}

\begin{figure}%  \begin{figure*}
\includegraphics[width=0.475\textwidth]{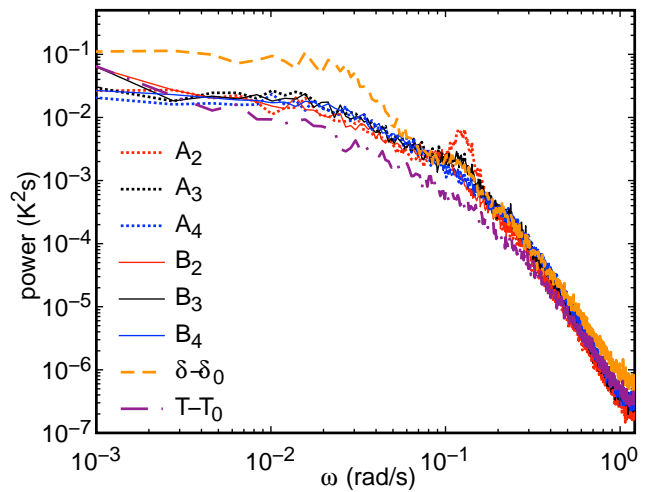}      
\caption{(color online) Power spectra of moments of the temperature profile as indicated in the legend, for a cylindrical container \cite{BA09}.  $A_2$ is the dominant oscillating mode, corresponding to a sloshing back and forth of the LSC.  This lacks the strong $A_3$ and $B_3$ moments found for a cubic container. 
}
\label{fig:fouriermomentscylinder} 
\end{figure} %\end{figure*}

%051023
To identify which moments of the oscillations observed in Sec.~\ref{sec:cube_structure} are depend on container geometry, we compare to data from a cylindrical cell at Ra$=1.1\times10^{10}$, $Pr=4.4$, and aspect ratio 1 \cite{BA09}.  Power spectra of the $A_n$ moments were reported in \cite{BA09}, and  here we additionally report $B_n$, $\delta$ and $T_0 -\langle T_0\rangle$ for that dataset.  Power spectra were calculated according to Eq.~\ref{eqn:powerspec}, smoothed by averaging over a range of $2.15\times10^{-3}$ rad/s, and shown in Fig.~\ref{fig:fouriermomentscylinder}. $A_2$ has the largest peak, in contrast with the cubic cell data in Fig.~\ref{fig:power_spec_momentsA} where $A_2$ was the only term that did not oscillate at the mid-height.  The oscillation in $A_2$ in the cylinder at mid-height is known as the sloshing aspect of the $n=1$ advected oscillation mode \cite{BA09}.   $B_3$ and $\delta-\langle\delta\rangle$ may also exhibit weak oscillations in Fig.~\ref{fig:fouriermomentscylinder}, but these are near our resolution limit.  The difference in oscillating moments between the cylinder in Fig.~\ref{fig:fouriermomentscylinder}  and in the cube in Figs.~\ref{fig:power_spec_momentsA} and \ref{fig:power_spec_momentsB} suggests that these new moments of oscillation of the temperature profile depend on container geometry; specifically the dominant $A_3$ moment followed by the $B_3$ moment.

\section{Model of the temperature profile}
\label{sec:model}

  \begin{figure}
\centerline{\includegraphics[width=0.475 \textwidth]{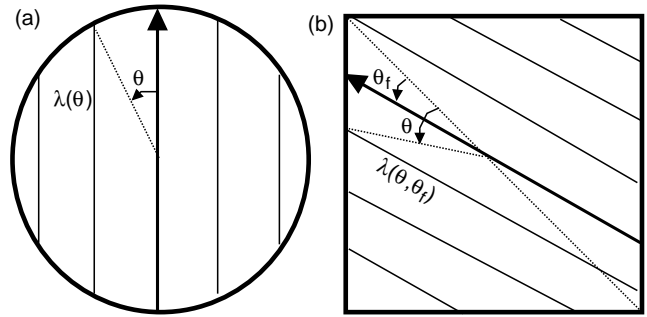}}
\caption{Pathlength of the LSC along the top and bottom plates as a function of the orientation $\theta_f$ of the flow  and coordinate $\theta$ for (a) a circular cross-section cell, and (b) a square cross-section cell.  The thick solid lines with arrows indicate the central path of the LSC.  The parallel lines indicate pathlengths at different $\theta$.
}  
\label{fig:pathlength_geometry}  
\end{figure}

%model assumptions
To understand the geometry-dependence of the oscillation of the shape of the temperature profile, we propose a minimal model.  Since the LSC transfers heat to and from the thermal boundary layers near the top and bottom plates, we hypothesize that the temperature difference $T-T_0$ across any streamline of the LSC is proportional to the pathlength $\lambda$ of that streamline along the top and bottom plates.  This is the result of a simplistic approximation of the heat conduction  if the boundary layer and flow are uniform along the plates, and the temperature difference between the boundary layer and the bulk is large compared to $T-T_0$ in the LSC.  To identify these streamlines we have to imagine an average over a time and length scale large enough to remove fluctuations, but short enough that the LSC orientation has not changed much.  The pathlength $\lambda$ can in general be a function of both the orientation $\theta_f$ of the flow at the edges of the top and bottom plates, and can vary around the cell with the coordinate $\theta$ at the wall of the container where measurements are made.   $\theta_f$ is not necessarily the same as the LSC orientation $\theta_0$ from the temperature profile fit, although they are expected to be closely related.  The pathlength as a function of $\theta$ depends on $\theta_0$, but the slosh displacement only shifts the nominal central plane of the LSC, not the pathlength as a function of $\theta$, so the pathlengths can be completely expressed in terms of $\theta_0$ and $\theta$.  We make the approximation that  the temperature profile obtained from the pathlength $\lambda$ is advected vertically and retains its shape, with the hot side of the LSC ($|\theta|\le \pi/2$) advected upward from the bottom plate, and the cold side of the cell advected downward from the top plate.   This ignores heat transfer by conduction in the bulk or turbulent mixing, as that is expected to smooth temperature profiles rather than produce higher order moments in the temperature profile.  Thus this model should not be used to characterize the dependence of the mean temperature profile on height. We also ignore any modification to the heat transport by counter-rolls in corners, as we do not know the flow fields in a cubic cell, so we do not know how big an effect they would have, how it depends on $\theta$, or how much the pathlength of corner rolls should count toward the sidewall temperature profile.  We approximate streamlines as straight across the plates, although even  the idealized traveling wave solutions of the advected oscillation mode would have curved streamlines \cite{BA09, JB20a}, resulting in a slight underestimate of the pathlength by using straight lines.  

With the above assumptions, the temperature profile is predicted to be 
\begin{equation}
T-T_0 \propto \frac{\lambda(\theta, \theta_f)}{H}sign[\cos(\theta-\theta_f)] \ ,
\label{eqn:tempprofile_model}
\end{equation}
where $sign[\cos(\theta-\theta_f)]$ is a step function to account for whether heat is  gained ($+1$) or lost ($-1$) from the hot or cold plates, respectively.  The pathlength $\lambda$ can be calculated based on the geometry of the container, as shown for example in Fig.~\ref{fig:pathlength_geometry} for circular and square cross sections. 
%cylinder
The predicted temperature profile reduces to a simple known result in the case of a circular cross section container:  $\lambda(\theta)/H = |\cos(\theta-\theta_f)|$, so Eq.~\ref{eqn:tempprofile_model} yields $T-T_0 \propto \cos(\theta-\theta_f)$, which  matches the shape of the measured mean temperature profile in cylindrical containers within 5\% \cite{BA07b}, if in this case the proportionality in Eq.~\ref{eqn:tempprofile_model} equals $\delta$,  and $\theta_f=\theta_0$.  With cylindrical symmetry, the pathlength does not depend on where $\theta=0$ is defined, and the flow orientation matches the extrema of the temperature profile ($\theta_f=\theta_0$). 
% $B_3$ is also the mode in which there is a slight perturbation to the mean temperature profile in the cylindrical cell with magnitude $0.05\langle\delta\rangle$ \cite{BA07b}.

%model for amplitude delta
The amplitude $\delta$ of the temperature profile can be predicted from a simple extension of the approximate rate of temperature change $\dot\delta$  of a parcel of fluid in the LSC due to conduction through the thermal boundary layer \cite{BA08a}, times the duration that the parcel of fluid advects past each plate $\mathcal{T}/4$ ($\mathcal{T} = 141$ s \cite{JB20a} is the LSC turnover time).  This yields the prediction $\delta=3\pi\kappa\Delta T \mathcal{N}\mathcal{T}/16L^2 = 0.8$ K ($\mathcal{N}=86$ \cite{FBNA05} is the Nusselt number), about twice the measured value, which is consistent within the level of approximation of the model of Brown \& Ahlers \cite{BA08a}.  This supports the assumption that the temperature profile can be quantitatively estimated based on heat conduction through the boundary layer.

\subsection{Predicted pathlength for a cubic cell}

\begin{figure}
\centerline{\includegraphics[width=0.475 \textwidth]{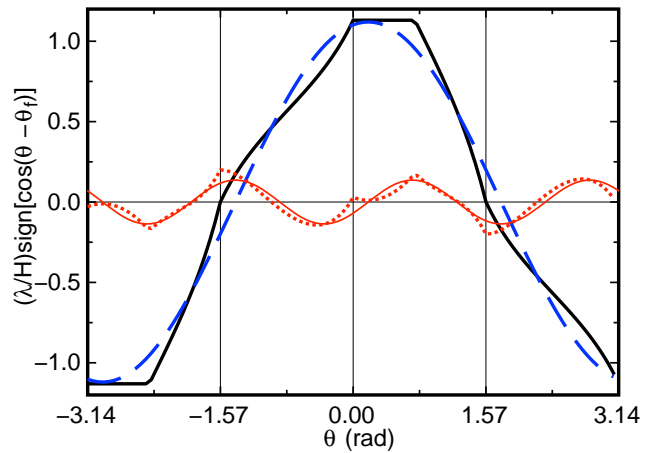}}
\caption{(color online) Thick solid black line: Example of a predicted temperature profile based on the pathlength $\lambda(\theta)/H$ along the top and bottom plates of a cubic cell for $\theta_f=0.3$ rad.    Dashed blue line: cosine fit (Eq.~\ref{eqn:tempprofile_model}) as a 1st-order approximation of the temperature profile.  Dotted red line:  difference between the predicted temperature profile and 1st-order cosine term.  Thin solid red line:  Fit of 3rd order sine moment ($A_3$) to the difference.  The difference is dominated by the $A_3$ moment.
}  
\label{fig:pathlength}
\end{figure}

%cube pathlength considerations
In a cell without circular symmetry, changes in the orientation $\theta_0$ lead to a change in the pathlength with both $\theta$ and $\theta_f$ as the shape of the container changes in the reference frame of the LSC.  The $n=1$ advected oscillation mode was observed in experiments with the same cubic cell that we report data from in this paper \cite{JB20a}.  The preferred flow orientation is aligned along a diagonal, with oscillations in $\theta_0$ centered on a diagonal. %Again, the slosh angle does not affect pathlengths at a particular $\theta$.   
While $\theta_0=0$ at the center of the top and bottom plates in the $n=1$ mode, assuming the LSC runs in a rectangular path along the inner wall of the container, the peak amplitude of the oscillation in $\theta_0$ at the edges of the plates is still expected to be $\sin[\pi/2\times\sqrt{2}/(1+\sqrt{2})]=0.79$ times the amplitude at the mid-height.  These hot and cold spots advect upward and downward, respectively, so that the thermal profile obtained at the top and bottom plates passes the mid-height at the same time.  The hot and cold spots are also predicted to advect azimuthally \cite{BA09}, and azimuthal flow has been observed \cite{VFKSSV19}.  Because the displacements in $\theta_0$ are in phase at the top and bottom plates for the $n=1$ mode, the restoring force pushes them in the same direction such that they shift the entire temperature profile so that $\theta_f$ is closer to zero at the mid-height. 
%a quarter-period later, so a quarter period behind the peak amplitude of the $\theta_0$-oscillation.  
Thus, even with this azimuthal advection, the predicted temperature profile at the mid-height is the same as the pathlength across the plate in terms of  $T(\theta-\theta_0$).

%pathlength  in cube
  For a square cross-section with origin $\theta=0$ in a corner as shown in Fig.~\ref{fig:pathlength_geometry}, the pathlength at the top and bottom plates is given by 
\begin{equation}	
\frac{\lambda(\theta)}{H} = \frac{1}{\cos(\pi/4-|\theta_f|)}
\label{eqn:pathlength_cube}
\end{equation}
for  $0<\theta<\pi/2$ and $[1+\tan(\theta-\pi/4)]/2 > \tan(\pi/4-|\theta_f|)$, which is a constant independent of $\theta$, and
\begin{equation}
\frac{\lambda(\theta)}{H} = \frac{1-\tan(\theta-\pi/4)}{2\sin(\pi/4-|\theta_f|)}
\label{eqn:pathlength_cube2}
\end{equation}
for  $0<\theta<\pi/2$ and $[1+\tan(\theta-\pi/4)]/2 <\tan(\pi/4-|\theta_f|)$, and 
\begin{equation}
\frac{\lambda(\theta)}{H} = \frac{1+\tan(\theta-\pi/4)}{2\cos(\pi/4-|\theta_f|)} \ .
\label{eqn:pathlength_cube3}
\end{equation}
for  $-\pi/2 < \theta < 0$.  This profile is repeated periodically every $\pi$ rad.   An example plot is shown in Fig.~\ref{fig:pathlength}.
%see notebook page 80 for derivation

%normalizing delta
To determine the proportionality coefficient in the temperature profile of Eq.~\ref{eqn:tempprofile_model} for a non-circular cross-section, we use the assumption of the Brown-Ahlers model that the flow velocity at the sidewall in the direction of the LSC (and thus $\delta)$ does not vary with small changes in diameter \cite{BA08b}. %Eq.~27. 
This choice had a significant consequence in determining the form of the potential in $\theta_0$, for example with potential minima at the corners \cite{BA08b}.
%$V_g(\theta_0)$, resulting in the scaling  $V_g(\theta_0)\sim D(\theta_0)^{-2}$ \cite{BA08b}.  
In contrast, allowing $\delta$ to vary with the pathlength along $\theta_0$ would have led to a different potential, %$V_g(\theta_0)\sim D(\theta_0)^{2}$
with a prediction that corners would no longer be the most preferred orientations, which disagrees with observations \cite{Zimin1978,ZML90,QX98, VPCG07, LE09, BJB16, FNAS17, GKKS18, VSFBFBR16, VFKSSV19,  JB20a}.  To keep consistent with the previous model assumption \cite{BA08b} and experimental observations, the  temperature profiles should be normalized by $\delta$ to remove variations in $\delta$ with $\theta_0$, so the proportionality coefficient should be $\langle\delta\rangle^2/\delta$, to retain the result that the mean value of the coefficient is $\langle\delta\rangle$.   However, since $\delta$ is calculated from the temperature profile, this normalization is done after the shape of the temperature profile is calculated and $\delta$ obtained from a fit of the temperature profile.  In practice, this leads to a correction of less than 0.3\% in the rms amplitudes of Fourier moments in the range of measured parameters shown in the following Figs.~\ref{fig:momentamplitudes_thetarms} and \ref{fig:moments_offset}, but more importantly it eliminates a large oscillation in $\delta$ that would occur if we instead assumed the proportionality in Eq.~\ref{eqn:tempprofile_model} were a constant.

\subsection{Predicted oscillating moments for a square cross-section}

%cube temp profile
In principle, oscillations in the shape of the temperature profile can occur if the pathlength of the LSC oscillates over time. 
Since we can analyze oscillation modes without adding noise, which would just obfuscate the results,   we include the effect of the oscillation in $\theta_0$ as a deterministic time series $\theta_f(t) = \theta_{f,max}\cos(\omega t)$, where $\theta_{f,max}$ is the amplitude of the oscillation in $\theta_f$ at the edge of the top and bottom plates.

  \begin{figure}
\centerline{\includegraphics[width=0.475 \textwidth]{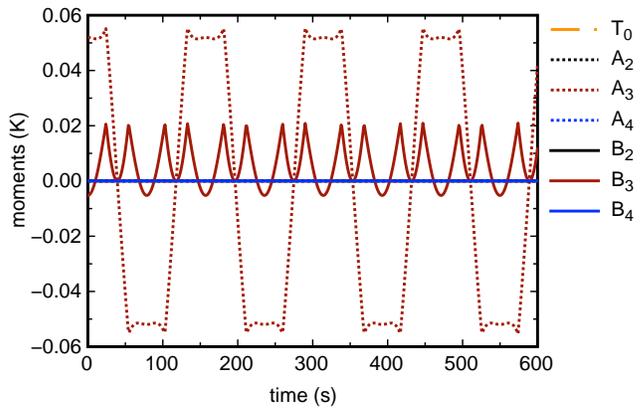}}
\caption{(color online) Time  series of Fourier moments of the temperature profile predicted from the assumed pathlength in Eqs.~\ref{eqn:pathlength_cube}, \ref{eqn:pathlength_cube2}, \ref{eqn:pathlength_cube3}.   The $A_3$ moment is dominant, followed by $B_3$.  The other moments shown in the legend are not excited. 
}  
\label{fig:moments_time_cubemodel}
\end{figure}

We calculate time series for the temperature profile from Eq.~\ref{eqn:tempprofile_model} using the pathlength of Eqs.~\ref{eqn:pathlength_cube}, \ref{eqn:pathlength_cube2}, and \ref{eqn:pathlength_cube3} for a square cross-section and oscillatory $\theta_f(t)$. % and $A_{2,i}(t)$.  
The temperature was sampled at 8 equally spaced orientations, mimicking our thermistor measurements.  The resulting temperature profile was fit at each time step to obtain the LSC orientation $\theta_0$ using Eq.~\ref{eqn:delta}. Fourier moments were then calculated at each time step using Eqs.~\ref{eqn:a_n} and \ref{eqn:b_n}.  An example time series of Fourier moments is shown in Fig.~\ref{fig:moments_time_cubemodel} for  $\langle\delta\rangle=0.431$ K and $\theta_{f,max}=0.3$ rad.  The $A_3$ moment has a dominant oscillation, followed by the $B_3$ moment.   The $A_3$ moment also has the dominant oscillation in the experimental data, followed by the $B_3$ moment (not counting  $A_4$ which violates symmetry expectations).

   \begin{figure}
\centerline{\includegraphics[width=0.475 \textwidth]{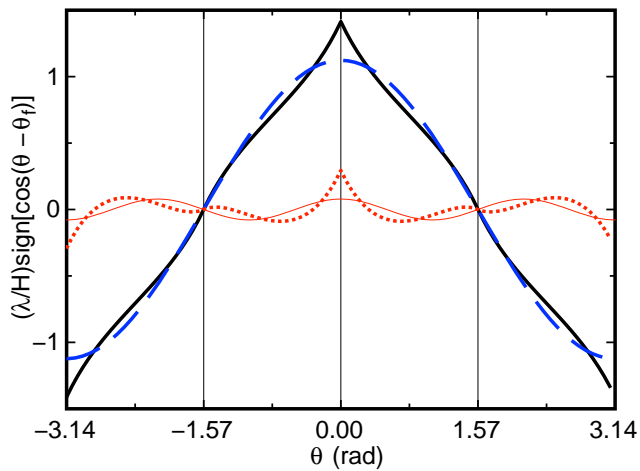}}
\caption{(color online) Thick solid black line: Example of the predicted temperature profile at an intermediate phase of the oscillation where  $\theta_f=0$ rad.    Dashed blue line: cosine fit (Eq.~\ref{eqn:tempprofile_model}) used for 1st-order approximation of the temperature profile.  Dotted red line:  difference between the predicted temperature profile and 1st-order cosine term.  Thin solid red line:  Fit of moment $B_3$ to the difference.  The difference has the symmetry of odd-order $B_n$ moments.
}
\label{fig:pathlength_intermediatephase}       
\end{figure}

 %explanation of dominance of A3 mode
 The dominance of the $A_3$ moment in the temperature profile is illustrated by comparing the  predicted temperature profile (Eq.~\ref{eqn:tempprofile_model}) at the phase where $\theta_f$ peaks to the 1st-order approximation of the cosine profile (Eq.~\ref{eqn:delta}) in Fig.~\ref{fig:pathlength}.  The difference between the predicted temperature profile and 1st-order cosine profile is shown as the dotted line in Fig.~\ref{fig:pathlength}.  A fit of the $A_3$ moment to the difference in Fig.~\ref{fig:pathlength} shows that it is dominated by that moment.  If $\theta_f$ oscillates, then Eq.~\ref{eqn:tempprofile_model} leads to oscillations of $A_3$ as is shown in  Fig.~\ref{fig:moments_time_cubemodel}. Thus, the dominant $A_3$ moment can be understood as a  result of the change in pathlength in the reference frame of the LSC as it oscillates around the corner of a cubic cell.  

%B3
To visualize the $B_3$ moment,  the predicted temperature profile for the intermediate phase where $\theta_f=0$ is shown in Fig.~\ref{fig:pathlength_intermediatephase}. It has the symmetry of odd-order $B_n$ moments (antisymmetric around $\theta=\pm \pi/2$ and symmetric around $\theta=0$).  With 8 thermistors, $B_3$ is the only odd-order $B_n$ moment that we can detect (shown by the fit in Fig.~\ref{fig:pathlength_intermediatephase}) due to the Nyquist limit.  The $B_3$ moment does not capture the full difference between the predicted temperature profile and the cosine fit shown in Fig.~\ref{fig:pathlength_intermediatephase}, and a $B_5$ moment with similar magnitude to the $B_3$ moment is apparent in the predicted profile by inspection.

%therm location effects
The details of the time series shown in Fig.~\ref{fig:moments_time_cubemodel} vary with the number and location of temperature measurements. For example, at the phase shown in Fig.~\ref{fig:pathlength_intermediatephase}, the arrangement of our thermistors yields a simulated measurement of $B_3=0$ at that phase even though the complete temperature profile in Fig.~\ref{fig:pathlength_intermediatephase} leads to a maximum of $B_3$ at this phase.  However, this moment is still detected in the time series at other intermediate phases, suggesting different thermistor arrangements will still detect the same moments, but with different amplitudes.  The  model temperature profile actually predicts $B_3$ is the strongest moment for small amplitude oscillations of $\theta_0$ (not shown), but our thermistors are not arranged to detect this moment optimally.  At least 10 evenly-spaced thermistors would be required to detect the predicted $B_5$ moment.
 
%difference between theta_0 and theta_f 
A feature of note is that the fit of the temperature profile in Fig.~\ref{fig:pathlength} yields the amplitude of the oscillation in $\theta_0$ to be $0.18$ rad instead of the input $\theta_{f,max}=0.3$ rad, indicating that $\theta_0$ from the cosine fit of the temperature profile underestimates the flow orientation $\theta_f$ in cubic cells.  $\theta_0$ and $\theta_f$ still oscillate together, and there is a one-to-one relationship between $\theta_0$ and $\theta_f$, so $\theta_0$ can work as a proxy for the flow direction as long as it is acknowledged that there is a conversion function between them in non-circular cross-section cells.

\subsection{Phase shifts}

Phase shifts between different signals can be identified from correlation functions, many of which were already reported for the cubic cell dataset \cite{JB20a}.
The observed in-phase behavior of $A_3$ at all 3 rows of thermistors is expected from the model, as it depends on the driving $\theta_0$-oscillation which is predicted to be in-phase at all 3 rows of thermistors in the $n=1$ advected mode \cite{JB20a} .   The observation that $A_3$ is $\pi$ rad out of phase with $\theta_0$ \cite{JB20a} is consistent with the picture that azimuthal advection shifts $\theta_f$ in Fig.~\ref{fig:pathlength} to 0 at the mid-height of the container so that $\theta_0$ is slightly negative while $A_3$ is positive.  This extra contribution to the measured $\theta_0$ superposed on the driving oscillation in $\theta_0$ may also explain why $\theta_0$ was not measured to be in-phase at all 3 rows of thermistors as predicted for the $n=1$ advected mode \cite{JB20a}.
 % not addressing whether A_3 is in phase with slosh or twist phase of oscillation

\subsection{Dependence on $\theta_{f,max}$}
\label{sec:parameter_dependence}

\begin{figure}
\centerline{\includegraphics[width=0.475 \textwidth]{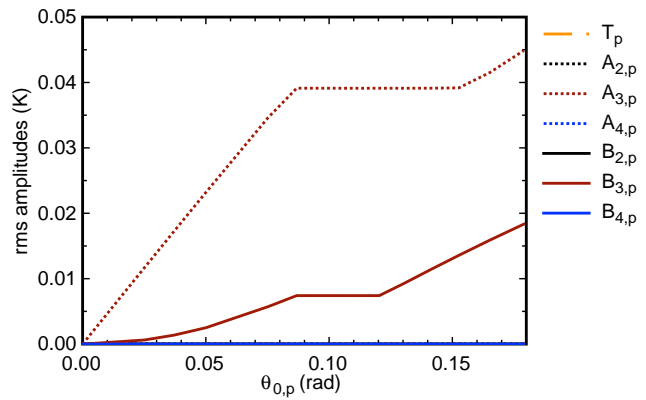}}
\caption{(color online) Predicted oscillation amplitudes of Fourier moments as a function of the amplitude of the driving oscillation amplitude $\theta_{0,p}$.
}  
\label{fig:momentamplitudes_thetarms}       
\end{figure}

%vary theta_max
To test whether the model quantitatively captures the observed moment amplitudes reported in Table \ref{tab:peak_power}, we varied the input oscillation amplitude $\theta_{f,max}$ and compare amplitudes of oscillating signals. Figure \ref{fig:momentamplitudes_thetarms} shows the root-mean-square (rms) amplitudes of the oscillating signals vs.~the rms of the oscillating component of the LSC orientation $\theta_{0,p}$ for $\langle\delta\rangle=0.431$ K (matching the measured value of $\langle\delta\rangle$).  Generally, $A_{3,p}$ is dominant, followed by $B_{3,p}$, and both moments increase nonlinearly with $\theta_{0,p}$.   We find an input value of $\theta_{f,max}=0.074$ rad  matches the observation of $\theta_{0,p}=0.037$ rad from Table \ref{tab:peak_power}.  For these input parameter values, the model yields the amplitudes $A_{3,p}=17$ mK and $B_{3,p}=1.4$ mK, somewhat smaller than measured root-mean-squared values of $A_{3,p}=28$ mK and $B_{3,p}=7$ mK for the middle row.   
%For small $\theta_{f,max} < 0.1$ rad, this corresponds to range where the lowest order expansion of sines and cosines in $\theta$ around a corner is good to better than 1\%.  

%smoothing
We tried smoothing the pathlength function as in Ref.~\cite{SBHT14}, and found that it did not introduce any different oscillating moments, and has less than 1\% effect on the magnitudes on moment amplitudes.

\subsection{Offset in $\theta_0$}

  \begin{figure}
\centerline{\includegraphics[width=0.475 \textwidth]{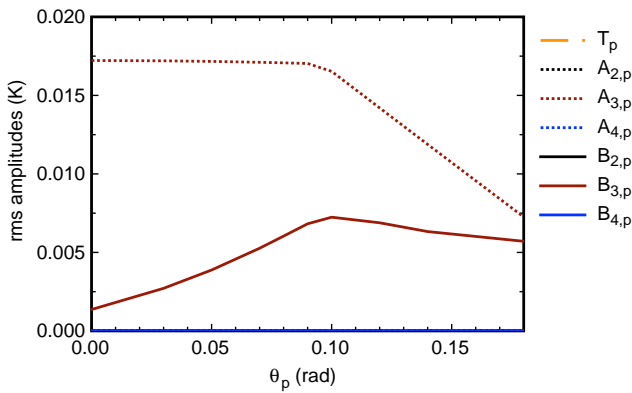}}
\caption{(color online) Predicted oscillation amplitudes of Fourier moments as a function of the offset of the preferred orientation $\theta_p$ from the corner for $\theta_{0,p}=0.037$ rad.  The offset of $\theta_p$ leads to an increase in $B_{3,p}$.
}  
\label{fig:moments_offset}       
\end{figure}

The preceding model assumes that the oscillation of $\theta_0$ is centered around the corner of the cell.  In actual experiments, the oscillation is not centered around the corner,  with the preferred orientation at an angle $\theta_p=0.05$ rad away from the corner, mainly due to a nonuniform temperature profile in the heating and cooling plates \cite{JB20a}.   This offset can change the induced temperature profile in the coordinate $\theta-\theta_0$, leading to different Fourier moment amplitudes than found in Sec.~\ref{sec:parameter_dependence}.
%In principle, it is possible that the nonuniform plate temperature profile, with standard deviation 90 mK in each plate, is large enough to induce oscillations of higher order Fourier moments in the temperature profile if there was an oscillation sweeping past different parts of the plate.  Can't measure $A_4$ moment since only have 4 thermistors in each plate.
 To determine the effect of this offset on the Fourier moments, we added the preferred orientation to the model for the oscillation of the flow direction: $\theta_f(t) = \theta_{f,max}\cos(\omega t)+\theta_p$. Figure \ref{fig:moments_offset} shows  rms amplitudes of Fourier moments as a function of the preferred orientation $\theta_p$ for $\theta_{0,p}=0.037$ rad. $B_{3,p}$ increases significantly with $\theta_p$, and no new moments are observed.  At the measured offset of $\theta_p=0.05$ rad, the model yields $A_{3,p}=17$ mK and $B_{3,p}=3.9$ mK.   This is within 50\% of the observed magnitudes of 28 mK and 7 mK, respectively.  
 %middle row $\theta_p$ is 0.05 rad away, average of 3 rows is 0.02 rad away from corner, and switching the flow direction in the plates leads to a change of 0.01 rad in $\theta_p$

\subsection{Predicted temperature profile of sloshing oscillations in a circular cross-section container}

 \begin{figure}
\centerline{\includegraphics[width=0.475 \textwidth]{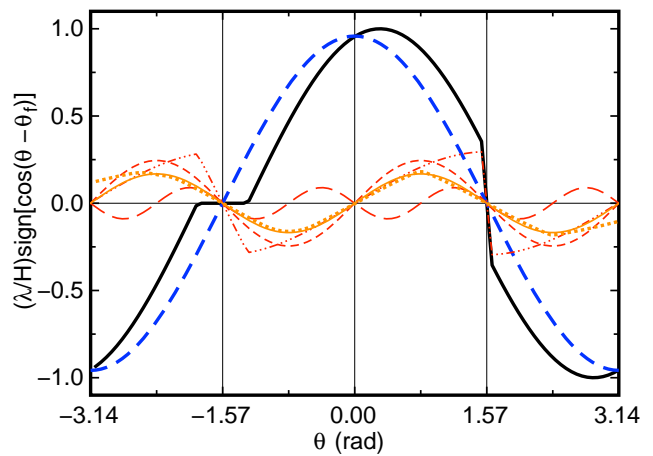}}
\caption{(color online) Thick solid black line: Example of predicted temperature profile due to the $n=2$ oscillation in a cylinder for $\theta_{f}=0.3$ rad.   Thick dashed blue line: cosine fit (Eq.~\ref{eqn:tempprofile_model}) as a 1st-order approximation of the temperature profile.  Dashed-dotted red line:  difference between the predicted temperature profile and 1st-order cosine term.  Short-dashed red line:  Fit of $A_2$ to the difference.  Long-dashed red line: Fit of $A_4$ to the difference.  Dotted orange line: difference between the predicted temperature profile and 1st-order cosine term smoothed over a range  $\pm\pi/4$.  Thin solid orange line: fit of the $A_2$ moment to the smoothed difference.  In either case, the difference is dominated by the $A_2$ moment, as observed in experiments.  The $A_4$ moment is highly dependent on smoothing.
}
\label{fig:pathlengthcylslosh}       
\end{figure}

Since the pathlength model can predict the existence of the new  $A_3$ and $B_3$ moments, it is worth checking if it can recover the temperature profiles of other known oscillations in temperature profiles, in particular the sloshing oscillation in a circular cross-section container.   The observed sloshing mode has peak amplitude at the mid-height \cite{XZZCX09, ZXZSX09, BA09}, and an oscillating $A_2$ moment \cite{BA09}.  This is part of the $n=2$ advected oscillation mode (with 2 oscillation periods per LSC rotation) which also has a twisting component corresponding to oscillations in $\theta_0$ at the top and bottom rows of thermistors out of phase with each other \cite{FA04, BA09}.  The twisting oscillation is $\pi/2$ rad out of phase with the sloshing oscillation.

%model
The pathlength-based model predicts that the cosine-shaped pathlengths on the hot and cold sides of the cell are advected upward and downward, respectively, to meet at the middle row.   Both cosine profiles at the hot and cold sides are advected azimuthally by a displacement $\theta_{f,max}$, but for the $n=1$ mode they advect in opposite directions in $\theta$ due to the twist displacements being on opposite sides at the hot and cold plates.  This results in a temperature profile where the hot and cold extrema are pushed closer together with the antisymmetry of the sloshing mode.  

%example
We show an example of this temperature profile in Fig.~\ref{fig:pathlengthcylslosh} for $\theta_{f}=0.3$ rad.  At the interfaces where the hot and cold sides are pushed together or pulled apart, we do not know how the temperature profile evolves, so as a simplest model we assume there is no advection past $\theta=\pi/2$ and the temperature around $-\pi/2$ is close to the average $T_0$.  The non-zero moments of the temperature profile for $n\le 4$ are $A_{2,p}\approx0.56\langle\delta\rangle\theta_{f,max}$  and $A_{4,p}\approx0.21\langle\delta\rangle\theta_{f,max}$.  The difference between  the predicted temperature profile and its 1st-order cosine fit, as well as the $A_2$ and $A_4$ moments, are shown in Fig.~\ref{fig:pathlengthcylslosh}.  This produces a dominant $A_2$ moment as observed, but also a significant $A_4$ moment, which is not observed in experiments \cite{BA09}.

%smoothing
Since a non-smooth temperature profile is unrealistic, as there will be some turbulent mixing to smooth the temperature profile, we considered the effects of smoothing in $\theta$.   Smoothing in general will have a tendency to reduce the contribution of higher order moments in the temperature profile.  For example, smoothing the difference in the predicted temperature profile from the 1st-order cosine fit over a range of $\pm \pi/4$ rad is shown in Fig.~\ref{fig:pathlengthcylslosh}.  This smoothing is even more dominated by the $A_2$ moment with $A_{2,p} \approx 0.39\langle\delta\rangle\theta_{f,max}$ and  $A_{4,p}\approx 0.04\langle\delta\rangle\theta_{f,max}$.  Due to the approximate nature of the model, we can only predict orders of magnitude for the effect of thermal mixing.  Since vertical temperature gradients are comparable to horizontal temperature gradients in the bulk \cite{BA07b}, the turbulent thermal mixing which is solely responsible for the vertical temperature gradient is comparable in magnitude to effects that generate the azimuthal temperature profile, and thus we cannot give a precise prediction of how much smoothing is expected from mixing.  Even without a precise prediction of the effects of thermal mixing, we can still say that the temperature profile is a combination of $A_2$ and potentially $A_4$ terms.  Note that including the effects of thermal mixing more generally in the model  is expected to similarly smooth the temperature profile, reducing the magnitudes of higher order moments more severely than lower order moments, but would not fundamentally change which moments can be excited by the geometry, or the order of magnitude estimates.

%Note that the advected wave solution where the slosh displacement and $\theta_0$ correspond to $A_2,p =0.5 \langle\delta\rangle\theta_{f,max}/\sqrt{2}$ in the small-angle limit if the deviation from the sinusoidal temperature profile only comes from the $A_2$ moment.  

%magnitudes
To compare the magnitudes of the predicted moments with experimental measurements, we measure an amplitude from the peak power of $\theta_{b,p} = 0.094$ rad at the bottom row of thermistors to represent the oscillation in $\theta_0$ for data from the cylindrical container.  The amplitude of the advected oscillation in $\theta_0$ at the bottom row of thermistors is not as large as at the edge of the bottom plate.  If we assume a rectangular pathlength around the inner edge of the aspect ratio 1 container, then $\theta_{f,max}= \theta_{b,p}\sin(\pi/4)/\sin(\pi/8) = 0.173$ rad.  For this $\theta_{f,max}$ and the observed $\langle\delta\rangle=0.255$ K, Eq.~\ref{eqn:tempprofile_model} predicts $A_{2,p} =  17-24$ mK and $A_4= 2-9$ mK, where the range is given from zero smoothing to smoothing by $\pm \pi/2$.  This overestimates the measured $A_{2,bot}=11$ mK at the mid-height in the cylindrical container by about a factor of 2, and is 1-4 times the resolution limit for measurements of the $A_4$ moment, so is reasonably consistent with the measurement of $A_2$ and lack of observed $A_4$ for an approximate model.
%LSC width $pi/10$  for the purpose of forcing, but may not be same for temperature profile smoothing
%\langle\delta\rangle=0.255 K
%background power is 67\% of peak power-bg for alpha. 15 mK without subtracting background. 
%background is 5 times larger than peak for theta at bottom row
% alternate option: location of hot and cold plume emission at plates oscillates and is advected upward, rather than horizontal advection.  In this case, the slosh profile does not come from the geometry.

%temperature profile of n=1 slosh mode
The situation is expected to be similar for the sloshing aspect of the  $n=1$ advected oscillation mode in non-circular cross sections.  The $n=1$ mode includes  sloshing at the top and bottom rows of thermistors, but not at the mid-height \cite{BA09,JB20a}.  We expect horizontal advection of the temperature profile will  move  $\theta_f$ from the in-phase twist displacement near the top and bottom plates by different amounts due to the different times it takes to reach the top and bottom rows of thermistors from the top and bottom plates, so that the two temperature extrema have moved closer together to produce a slosh displacement.  Due to the conceptual similarity with the slosh profile for the cylinder, it is not shown here.  Due to the ambiguity of how to define the temperature profile near the border of the hot and cold sides, any breakdown of the slosh profile into $A_2$ and $A_4$ components would be again highly dependent on assumptions about smoothing, and our measurement of $A_4$ is plagued by large asymmetries of the setup, any quantitative results we could show would not be very informative.  

%jumprope
The jumprope oscillation is the other mode that is known to cause an oscillation in the shape of the temperature profile, specifically with the symmetries of even order $B_n$ moments and $T_0$ \cite{VHGA18}.  However, since this is known to consist of the core of the LSC oscillating around in the central plane of the LSC, it does not involve changes in pathlengths of parallel paths along the top and bottom plates in the sense discussed here, so that temperature profile oscillation should be described by a different mechanism.

\section{Summary}

%cubic observation
We observe an oscillation in the shape of the LSC  expressed in terms of Fourier moments of the azimuthal temperature profile relative to the LSC orientation $\theta_0$.  In a cubic container, the oscillation is dominated by the 3rd-order sine moment ($A_3$), followed by the 3rd-order cosine moment ($B_3)$ (Figs.~\ref{fig:power_spec_momentsA}, \ref{fig:power_spec_momentsB}).  This oscillation in the temperature profile lacks the symmetry in $\theta_0$ of the jumprope mode, lacks the antisymmetry in $\theta_0$ of the sloshing mode, and was not found in cylindrical containers like those other modes, making it distinct from those known oscillation modes.  

%explanation
An explanation for this new oscillation in the temperature profile is that the heat conducted from the thermal boundary layers to each streamline of the LSC is proportional to the pathlength of the streamlines along  the top and bottom plates.   In a non-circular-cross-section cell, the oscillation of the LSC orientation $\theta_0$ results in the pathlength as a function of $\theta-\theta_0$ oscillating in the reference frame of the LSC, and thus oscillations in the temperature profile.  The detailed collection of moments can be estimated based on the container geometry (Fig.~\ref{fig:pathlength_geometry}).
%using Eq.~\ref{eqn:tempprofile_model} and pathlengths as in Eqs.~\ref{eqn:pathlength_cube}, \ref{eqn:pathlength_cube2}, and \ref{eqn:pathlength_cube3} 
In a cubic cell, this minimal model predicts a dominant $A_3$ moment followed by $B_3$ (Figs.~\ref{fig:pathlength}, \ref{fig:moments_time_cubemodel}, \ref{fig:pathlength_intermediatephase}), with magnitudes about 50\% larger than measured values when using experimental input for the advected oscillation amplitude in $\theta_0$ (Figs.~\ref{fig:momentamplitudes_thetarms}, \ref{fig:moments_offset}).    Thus, the oscillation in the temperature profile in a non-circular cross section is not an entirely new oscillation mode, but a modification of the temperature profile  due to previously known oscillations of the LSC orientation $\theta_0$.

%circular
A circular-cross-section cell is special in that the pathlength along the plate is independent of $\theta_0$, and so no oscillations in the shape of the temperature profile are induced as $\theta_0$ oscillates.  In a cylindrical cell, the model produces the sinusoidal mean temperature profile with a sloshing oscillation dominated by the 2nd order sine moment ($A_2$) (Fig.~\ref{fig:pathlengthcylslosh}), in agreement with observations in that geometry (Fig.~\ref{fig:fouriermomentscylinder})\cite{BA07b, BA09}. 

%A4
We also show that observation of oscillations of $A_4$ in phase with the oscillation of $\theta_0$ break the symmetry between the hot and cold sides of the LSC. These are likely a consequence of asymmetries of the setup, or possibly a long-lived symmetry-breaking mode.

 %big picture
%This work explains the oscillating Fourier moments of the temperature profile in a cubic cell that could not be explained by the advected oscillation model of Brown \& Ahlers \cite{JB20a}. 
This work demonstrates a low-dimensional approximate model can explain the qualitative shape and order of magnitude of the LSC temperature profile and some of the major perturbations to it in Rayleigh-B{\'e}nard convection.  It remains to be seen if quantitative predictions can be refined using more detailed information of the temperature and flow fields, for example to quantify the effects of corner-rolls on the temperature profile, or using local properties of the thermal boundary layer to calculate heat transport to the LSC.  It also remains to be seen if the basic concept of estimating heat transport from conduction along the boundary layer pathlength can inform modeling of large-scale coherent flow structures in other systems.

%jumprope-maybe. 
%corner rolls -maybe

\section{Acknowledgements}

We thank the University of California, Santa Barbara machine shop and K. Faysal for helping with construction of the experimental apparatus.  This work was supported by Grant CBET-1255541 of the U.S. National Science Foundation.  

\section{Author Contributions}

Experimental measurements and data analysis for the cubic cell was carried out by D. Ji.  Modeling, and analysis of data from the cylindrical cell was done by E. Brown.  The draft was written by E. Brown.

%\bibliography{../rbc}


\end{document}